\documentclass[twocolumn,showpacs,preprintnumbers,amsmath,amssymb,prl,superscriptaddress]{revtex4}

\usepackage{graphics}

\newcommand{\abs}[1]{\left\vert #1\right\vert}

\begin{document}

\title{
Spin current generation and detection by a double quantum dot
structure}
\author{J.~P.~Dahlhaus$^1$, S.~Maier$^1$, and A.~Komnik}
\affiliation{Institut f\"ur Theoretische Physik, Universit\"at Heidelberg,\\
 Philosophenweg 19, D-69120 Heidelberg, Germany}
\date{\today}

\begin{abstract}
We propose a device acting as a spin valve which is based on a
double quantum dot structure with parallel topology. Using the
exact analytical solution for the noninteracting case we argue
that, at a certain constellation of system parameters and
externally applied fields, the electric current through the
constriction can become almost fully spin-polarized. We discuss
the influence of the coupling asymmetry, finite temperatures and
interactions on the efficiency of the device and make predictions
for the experimental realization of the effect.
\end{abstract}

\pacs{73.23.-b, 73.63.Kv, 85.75.-d}

\maketitle

Future progress in the recently very fast developing field of
spintronics heavily depends on reliable techniques for the
generation and detection of spin-polarized electric currents
\cite{Wolf2001}. While the majority of proposed devices uses
ferromagnetic electrodes of some form, purely semiconductor-based
structures possess a number of advantages such as lower power
consumption and smaller dimensions as well as better integration
options into the conventional circuitry. Probably the best studied
elementary structures are quantum point contacts and quantum dots
which can, among other things, be used to induce spin currents. Up
to now numerous studies have been devoted to the investigation of
these possibilities, to name just a few of them:
\cite{PhysRevB.65.165303,pustilnik:201301,Tao2008}. In its
simplest form a quantum dot is just an isolated electronic energy
level coupled to a number of metallic electrodes by tunneling (and
possibly capacitively). The transmission coefficient is then of
Lorentzian shape with half-width $\Gamma$ given by the contact
transparency between the dot and the electrode. Its resonant
behavior immediately suggests one possibility for spin-polarized
current generation: the Zeeman-splitting of the level in a finite
external field leads to different transmission probabilities for
electrons with different spin orientation (the magnetic field is
assumed to be finite only on the dot). This method is, however,
extremely inefficient since the level-splitting is of the order
$0.025$ meV/T for GaAs-based heterostructures and thus even in
strong fields significantly smaller than the typical $\Gamma$
ranging between $0.1-10$ meV
\cite{Goldhaber-Gordon1998,Cronenwett1998,Schmid1998}. Generally,
the current through the constriction grows with increasing
$\Gamma$ such that a compromise must be arranged between the spin
polarization quality factor and the current strength. Therefore
one has to search for systems which show up transmission
properties with even higher degree of nonlinearity than that of a
simple (non-interacting) dot. Exactly this situation can be found
in double quantum dot systems \cite{RevModPhys.75.1}.

In general a double quantum dot structure even in its simplest
incarnation, in which it is modelled by two coupled Anderson
impurities, is described by a large number of parameters. The
corresponding Hamiltonian is given by \cite{PhysRevB.68.125326}
\begin{eqnarray}
 H = H_0 + H_I + H_T \, .
\end{eqnarray}
$H_0$ is the part describing the two dots ($i=1,2$) via respective
fermion annihilators/creators $d_{i,\sigma}^\dag, d_{i,\sigma}$
with spin variable $\sigma = \uparrow,\downarrow=\pm$ and two
 (left/right, $\alpha=L,R$) metallic electrodes. These are modelled by
 free fermionic
 continua with field operators $\psi_{\alpha,\sigma}(x)$, which are kept
 at chemical potentials $\mu_\alpha$,
 \begin{eqnarray}
  H_0 =\sum_{\alpha,\sigma}  H_\alpha[\psi_{\alpha,\sigma}] + \sum_\sigma\sum_{i=1,2}  (E_i + \mu_B g \sigma h/2)
  \, d^\dag_{i,\sigma} d_{i,\sigma} \, ,
 \end{eqnarray}
where $E_i$ are the bare dot level energies, $\mu_B$ is the Bohr's
magneton, $g$ and $h$ are the Land\'{e} factor and magnetic field,
respectively. Electron exchange between the electrodes and the
dots is accomplished by
\begin{eqnarray}
 H_T = \sum_{i, \alpha, \sigma}
 \gamma_{i, \alpha} \, d_{i,
 \sigma}^\dag \psi_{\alpha, \sigma}(0)   +
 \gamma_\perp \, d^\dag_{1,\sigma} d_{2,\sigma}  + \mbox{H.c.} \, ,
\end{eqnarray}
where $\gamma_{i, \alpha}$ is the tunneling amplitude between dot
$i$ and electrode $\alpha$ and $\gamma_\perp$ is responsible for
the electron exchange between the dots. The tunnelling is assumed
to be local and occur at $x=0$ in the coordinate system of the
respective electrode. In general, the tunneling amplitudes are
allowed to be complex. Finally, the interactions in the system are
taken into account via the last term,
\begin{eqnarray}
  H_I = \sum_i U_i \,  \, n_{i, \uparrow} \, n_{i, \downarrow}
  + \sum_{\sigma,\sigma'} U_\perp
   n_{1, \sigma} \,
    n_{2, \sigma'}\,
    .
\end{eqnarray}
where $n_{i, \sigma} = d^\dag_{i, \sigma}  d_{i, \sigma}$. While
$U_i$ is responsible for the intradot interaction, $U_{\perp}$
describes the interdot correlation.

\begin{figure}[h]
\includegraphics{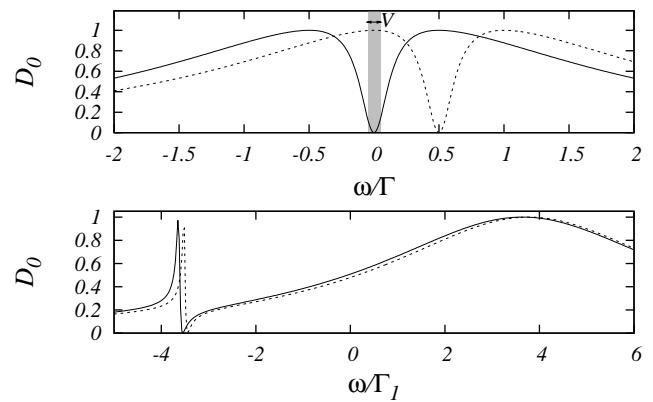}
\caption{\emph{Upper panel:} the operation mode of the spin
filter. The energy levels are set to $E_{1,2}=\pm 0.5$. The offset
$2h=0.5$ is chosen in such a way that the efficiency of the filter
reaches its maximum value. All energies are measured in units of
$\Gamma$. The shaded area represents a voltage window for the
generation of highly spin-polarized current. \emph{Lower panel:}
the effect of asymmetric coupling and inter-dot tunneling. The
Fano-lineshape can clearly be identified. Here, the parameters are
$E_{1,2}=\pm 1$, $2h=0.15$, $\Gamma_2 = 0.9$ and $\gamma_\perp =
3.5$. All energies are measured in units of $\Gamma_1$. The local
maximum of the dotted line coincides with the zero of the solid
line to achieve optimal operation of the spin filter.}
\label{fig:tcoeff}
\end{figure}

In order to illustrate our idea we first neglect the interactions,
use $\gamma_\perp =0$ and equalize all other tunneling amplitudes
to $\gamma$. Transport in such a setup has been investigated in
great detail in a number of works, see e.~g.
\cite{PhysRevB.47.6835,PhysRevLett.84.1035,PhysRevLett.87.256802,PhysRevB.65.045316}.
The fundamental result for the energy-dependent transmission
coefficient reads \cite{PhysRevB.65.245301}
\begin{eqnarray}             \label{coeff}
 D_0(\omega) = \frac{\Gamma^2}{\left[ 1/(\omega - E_1) + 1/(\omega
 - E_2)\right]^{-2} + \Gamma^2} \, ,
\end{eqnarray}
where $\Gamma = 2 \pi \rho_0 |\gamma|^2$ is the dot-lead contact
transparency with dimension of energy. It consists of the
tunneling amplitudes and the local density of states $\rho_0$ in
the leads which is assumed to be very weakly energy-dependent in
the relevant range of energies. When the energies of both dots are
equal the system is equivalent to the conventional single-site
Anderson model as far as the transmission properties are
concerned. However, contrary to a single-site quantum dot here
\emph{perfect reflection} is possible when the energy of the
incident particles is given by $\omega_0 = (E_1 + E_2)/2$ as soon
as $E_{1,2}$ become different. We speculate that this kind of
destructive interference is very similar to the one leading to
weak localization. Electrons with energy $\omega_0$ which travel
in the (anti)clockwise direction through the device (and thus on
the time-reversal equivalent paths) experience exactly the same
phase shift leading to constructive interference at the starting
point. The precise form of this kind of \emph{antiresonance} can
be found by rewriting the transmission coefficient (\ref{coeff})
in the following form
\begin{eqnarray}
\label{eq:lorenzians}
 D_0(\omega) = \frac{\Gamma}{\sqrt{\Gamma^2 - E^2}} \left(
 \frac{\Omega_+^2}{\omega^2 + \Omega_+^2} - \frac{\Omega_-^2}{\omega^2 +
 \Omega_-^2} \right) \, ,
\end{eqnarray}
where we measure the energy from $\omega_0$ and restrict ourselves
to $\Gamma> E$, $E=E_1=-E_2$. Unsurprisingly it is a difference of
two Lorentz-shaped curves with the widths $\Omega_\pm = \Gamma \pm
\sqrt{\Gamma^2 - E^2}$. The antiresonance thus can be made
extremely sharp by choosing $E$ very small in comparison to
$\Gamma$ by appropriate choice of the gate voltages. In presence
of the magnetic field $h$ (we assume that it is not generating any
Aharonov-Bohm phase either due to the small area of the dot
structure or due to its in-plane orientation) the antiresonance
splits in two for electrons with different spin orientation
$\sigma$,
\begin{eqnarray}
 D_\sigma (\omega) = D_0(\omega + \sigma h) \, ,
\end{eqnarray}
where we have redefined $h = \mu_B g h/2$ to become the effective
energy scale generated by the magnetic field. Thus the
transmission coefficients for the electrons with different spin
orientations are completely different, see Fig. ~\ref{fig:tcoeff}.
In fact, the electrons with the energy matching their `own'
antiresonance are perfectly reflected while the ones with opposite
spin orientation can be made to pass through the structure almost
unimpeded.
This is in strong contrast to the single dot structure discussed
in the introduction, where perfect transport suppression is
difficult to achieve.

Applying finite bias voltage $V$ across the double dot we can in
fact generate almost fully spin polarized current as long as $V
\ll \Omega_-$ and both chemical potentials are symmetrically
arranged around the preselected antiresonance, see
Fig.~\ref{fig:tcoeff}. In the experimental realization this
procedure would amount to a fine-tuning of the dot level energies
$E_{1,2}$ as well as of the applied magnetic field. The points
where the current is maximally spin-polarized coincide exactly
with the points, where the spin-unresolved (conventional)
transmission has a dip. Needless to say, in analogy to optical
polarizators this effect can also be used for detection of
spin-polarized currents.

There are different mechanisms which can destroy the interference
and thus significantly affect the quality of spin filtering: (i)
finite temperature effects; (ii) the finite interdot tunneling
amplitude $\gamma_\perp$ as well as the coupling asymmetry;  (iii)
the effects of intra- as well as interdot interactions.\footnote{
There are also extrinsic factors which
 e.~g. leakage currents or a
coupling of the dots to heat baths. Those can also be dealt with
in the same fashion.} While (i) and (ii) can be (at least in
principle) very well controlled in experiments the interactions
can be influenced only slightly.

\begin{figure}[h]
\includegraphics{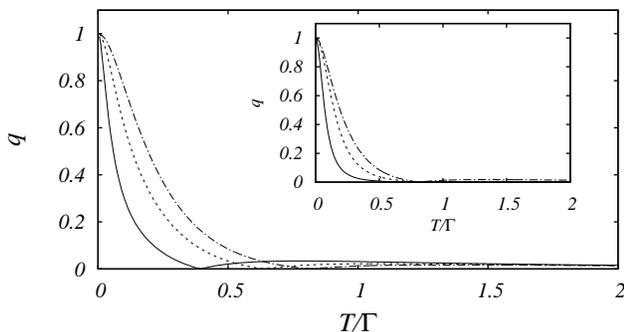}
\caption{Temperature dependence of the quality factor as defined
in Eq.~(\ref{quality}). \emph{Main plot}: as a function of the dot
energies
$E=(0.5,0.75,0.95)=(\text{solid},\text{dotted},\text{dot-dashed})$
for fixed $h=0.5$. \emph{Inset}: the same for $E=0.95$ and
different magnetic fields $h=(0.1, 0.25, 0.5)$. All energies are
measured in units of $\Gamma$. } \label{fig:qfactor1}
\end{figure}

We first analyze the finite-$T$ case. We assume the voltage
applied symmetrically around the $\sigma = \uparrow=+$
antiresonance, then the spin-resolved currents are given by
\begin{eqnarray}              \label{I_pm}
 I_{\pm} &=& G_0 \int d \omega D_0(\omega + h_\pm)
 \nonumber \\
 &\times&
 \left[ n_F(\omega -
 V/2) - n_F(\omega + V/2) \right] \, ,
\end{eqnarray}
where $n_F(\omega)=1/[\exp(\omega/T)+1]$ is the Fermi distribution
function, $G_0=e^2/h$ the conductance quantum per spin orientation
and the Zeeman splitting of the dot level energies is taken into
account by $h_+ = 0$ and $h_-=-2h$ (Without an additional gating
of the dot levels by $h$ the antiresonances would, of course, lie
at $\pm h$. In order to achieve optimal spin filtering we choose
to perform such adjustment of $E_{1,2}$). It is sensible to make
predictions for the universal linear response regime first, where
the linear conductance $G_\pm = I_\pm(V)/V$ at $V\rightarrow 0$ is
the fundamental quantity. Then for the quality factor of the spin
filtering we obtain the following result
\begin{eqnarray}           \label{quality}
 q = \left| \frac{G_+ - G_-}{G_+ + G_-} \right|
  \, ,
\end{eqnarray}
where $G_\pm$ is defined by
\begin{eqnarray}
G_\pm &=& \frac{\Gamma}{\sqrt{\Gamma^2-E^2}} \Biggl[\sum_{r,s,t=\pm}
\frac{rs t\Omega_r}{4\pi T} \psi'\left(\frac{1}{2}+t\frac{i h_\pm+s\Omega_r}
{2\pi T}\right) \nonumber \\
&+& \sum_{r,s=\pm}\frac{s \abs{\Omega_r}}{4\pi T} \psi'\left(\frac{1}{2}
-\frac{i s h_\pm+ \abs{\Omega_r}}{2\pi T}\right)\Biggr]
\end{eqnarray}
and $\psi'$ is the derivative of the digamma function. As a
function of temperature it is plotted in Fig.~\ref{fig:qfactor1}.
Well below $T=\Gamma$ very high quality factors are achievable.
The effect is more robust for higher values of bare dot energies
and applied local field $h$.

As far as the asymmetry and $\gamma_\perp$ issues are concerned
analytical results are possible as well. Interestingly, in both
cases the perfect destructive interference is still possible.
Hence the spin filter can still be realized as illustrated in the
lower panel of of Fig.~\ref{fig:tcoeff}.

\begin{figure}
\centering
\includegraphics{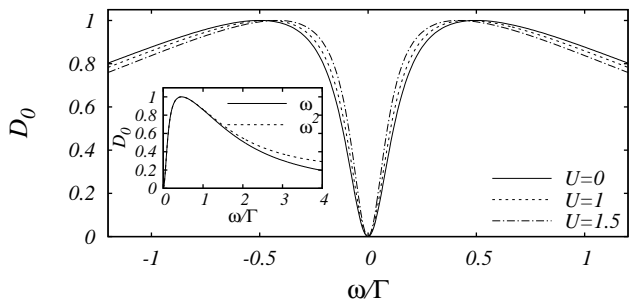}
\caption{Influence of weak interactions on the effective
tranmission coefficient around the antiresonance. \emph{Main
plot:} antiresonance with self-energy correction up to linear
order in $\omega$ and second order in $U/\pi$. The energy levels
of the dots are $E=0.5$. \emph{Inset} illustrates the effect of
second order corrections in $\omega$ and $U/\pi$. Again, the dot
energies are $E=0.5$ and the intra-dot interaction is $U=1$.
Energies are measured in units of $\Gamma$.} \label{fig:efftrans}
\end{figure}

\begin{figure}
\centering
\includegraphics{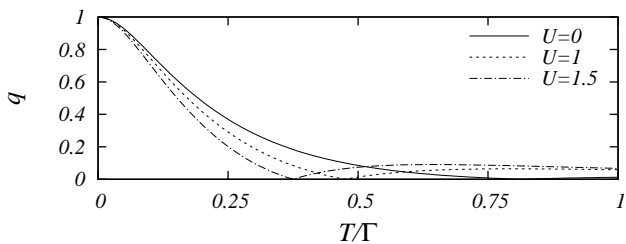}
\caption{Temperature dependence of the quality factor in the
interacting case for $E=0.95$ and $h=0.5$ and different
interaction strengths. Energies are measured in units of
$\Gamma$.} \label{fig:qfactoru}
\end{figure}

The electron-electron interactions on such small quantum dots can
usually be quite strong. Despite quite extensive literature the
fate of the antiresonance in the case of finite interactions has
not yet been addressed. We consider the onsite interactions with
amplitudes $U_{1,2}$ first. The transmission coefficient
(\ref{coeff}) can be found in different ways. One of them is the
straightforward calculation of the scattering amplitude of a
structure consisting of two Y-shaped junctions arranged in a ring
geometry with two in- and outputs between which the two dots are
arranged \cite{Buttiker1984,PhysRevB.67.205303}. It is the special
property of the dot scattering phases or transmission/reflection
amplitudes $r_\alpha,t_\alpha$ as functions of energy $\omega$,
which give rise to the antiresonance. In the noninteracting case
they  are given by\cite{PhysRevB.65.245301}
\begin{eqnarray}              \label{eqX}
 r_\alpha = \frac{- i \Gamma}{\omega - E_\alpha - i \Gamma} \, ,
 \, \, \, \, \,
% \nonumber \\
 t_\alpha = \frac{\omega - E_\alpha}{\omega - E_\alpha - i \Gamma}
 \, .
\end{eqnarray}
Being plugged into the expression for the full transmission of the
structure \cite{Buttiker1984}:
\begin{eqnarray}
D_0=4\abs{\frac{t_1 t_2\bar{r}_1-\bar{t}_1 t_2 r_2 - t_1 t_2
\bar{r}_1 +\bar{t}_1\bar{t}_2 r_2}{t_1 t_2-\bar{t}_1\bar{t}_2-\bar{t}_2
r_1-\bar{t}_1 r_2+\bar{t}_1 \bar{r}_2}}^2
\end{eqnarray}
they immediately lead to (\ref{coeff}). In fact, (\ref{eqX}) are
related to the retarded Green's function (GF) of the individual
dots \cite{Meir1992,Yamada1975a}:
\begin{eqnarray}
 G^R_\alpha(\omega) = \frac{1}{\omega - E_\alpha + i \Gamma} \, .
\end{eqnarray}
On the other hand, the retarded GF (or the transmission matrix)
for the case with finite interactions is known to possess the
representation \cite{Yamada1975a}
\begin{eqnarray}           \label{GF}
  G^R_\alpha(\omega) = \frac{1}{\omega - E_\alpha - \mbox{Re} \Sigma^R(\omega)
  + i [\Gamma - \mbox{Im} \Sigma^R(\omega) ]} \, ,
\end{eqnarray}
where $\Sigma^R(\omega)$ is the self-energy due to the onsite
interaction. Because we are only interested in the transmission
properties around $\omega =0$ it is sufficient to possess
information about the self-energy behavior around this point. A
good approximation for the self-energy  is the one of the ordinary
Anderson impurity model. Luckily, there is a low-energy expansion
for this $\Sigma^R(\omega)$ due to
\cite{Yamada1975,Yamada1975a,Yosida1975,Yosida1970,Zlati'c1980,Zlati'c1983}.
The main message is that the leading order expansion in $\omega$
is provided by the correction to the real part \cite{Oguri2001a},
\begin{eqnarray}                        \label{repart}
 \mbox{Re} \Sigma^R_{\alpha \sigma} (\omega) = \chi_c
 (E_\alpha + U/2) + \sigma h \chi_s
 \nonumber \\
 + \left( 1 - \frac{\chi_c - \chi_s}{2} \right) \, \omega
  + \dots \, ,
\end{eqnarray}
where $\chi_{c,s}$ are the static charge/spin susceptibilities and
are known for \emph{arbitrary} $U$ from the Bethe ansatz
calculations \cite{Zlatiifmmodeacutecelse'cfi1983}. $(E_\alpha +
U/2)$ plays the role of the electron-hole symmetry breaking field.
Thus we conclude that up to the finite shift $\delta E_\alpha =
\mbox{Re} \Sigma^R_{\alpha \sigma} (0)$ the antiresonance survives
and we expect the same quality of spin filtering is achievable.
The next question about the antiresonance width can only be
answered with the next order expansion in $\omega$ at hand. Since
the leading order for the imaginary part of the self-energy is
$\omega^2$ (which is not surprising since it is responsible for
the dissipative part and thus for inelastic processes) the form of
the antiresonance is dominated by the second term of
(\ref{repart}). Then the transmission is given by
\begin{eqnarray}  \nonumber
D_0=\frac{\Gamma^2}{\Gamma^2+\left(\frac{1}{\omega\frac{\chi_{1c}
-\chi_{1s}}{2}-\left(E_1+\delta E_1
\right)}+\frac{1}{\omega\frac{\chi_{2c}
-\chi_{2s}}{2}-\left(E_2+\delta E_2\right)}\right)^{-2}} \, .
\end{eqnarray}
In case of a small-$U$ expansion \cite{Zlati'c1980} one can
rewrite this again as a sum of two Lorentzians. Apart from a
rescaling of the $\Omega_\pm \rightarrow \Omega_\pm / \alpha$
where
$\alpha=\frac{U^2}{2\pi^2}\left(3-\frac{\pi^2}{4}+\left(\frac{25}{3}
-\frac{3\pi^2}{4}\right)\frac{E^2}{\Gamma^3}\right)$ one finds
equation (\ref{eq:lorenzians}). The effects of higher order terms
of $\omega$ can be understood using the $U$-expansion results of
\cite{Zlati'c1980}. To illustrate their influence on the
transmission we plotted the antiresonance upon inclusion of the
$\omega^2$ terms in Fig.~\ref{fig:efftrans}. The effect of
interactions on the quality factor of spin filtering is presented
in Fig.~\ref{fig:qfactoru}.

\begin{figure}
\centering
\includegraphics{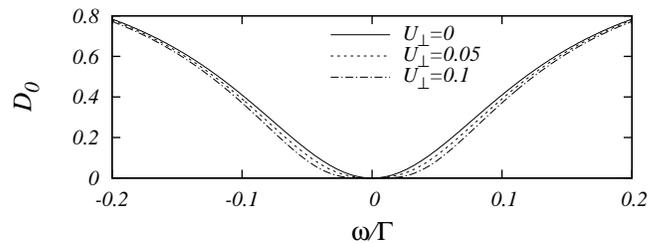}
\caption{Effect of inter-dot interactions on the transmission
coefficient around the antiresonance for different values of the
interaction strength. The energy levels of the dot are given by
$E=0.5\Gamma$. All energies are measured in units of $\Gamma$.}
\label{fig:transj}
\end{figure}

Interactions between electrons on different dots are often weak in
comparison to $U_{1,2}$. Thus we can treat them perturbatively. It
is a tedious but straightforward calculation so we suppress the
details. Already in the lowest order the correction to the
transmission coefficient reveals an interesting effect of
\emph{antiresonance enhancement}, see Fig.~\ref{fig:transj}. While
the intradot interactions appear to narrow the antiresonance the
effect of the interdot interactions is quite the opposite.

As we have shown above the perfect antiresonance is not destroyed
by the not too strong Coulomb interactions within the device.
However, we expect that this is not the case as soon as
interactions with the environment are included. These effects can
be discussed by modifying the respective dot Green's function
(\ref{GF}) or using the appropriate self-energy. It is not only
possible to analyze perturbations with particle exchange with the
environment (leakage currents etc.) but also to include
interactions with phonon baths and electromagnetic environments.

To conclude, we propose a device for spin-polarized current
generation and detection. It is based on the double quantum dot
structure and operates around the antiresonance in transmission
achieved at certain constellation of dot parameters and external
fields. We discuss the quality factor of spin filtering as well as
its robustness against intrinsic and extrinsic factors such as
finite temperature, interaction effects and contact to the
environment. We expect that the discussed spin-filtering
techniques can be implemented in the up-to-date double quantum dot
devices such as those presented in
\cite{PhysRevLett.87.256802,Wilhelm2002385}.

\acknowledgments The authors would like to thank T.~L.~Schmidt for
many interesting discussions. The financial support was provided
by the DFG under grant No.~KO~2235/2 and by the Kompetenznetz
``Funktionelle Nanostrukturen III'' of the Landesstiftung
Baden-W\"urttemberg (Germany).

\bibliography{DD_paper_bib}

\begin{thebibliography}{26}
\expandafter\ifx\csname natexlab\endcsname\relax\def\natexlab#1{#1}\fi
\expandafter\ifx\csname bibnamefont\endcsname\relax
  \def\bibnamefont#1{#1}\fi
\expandafter\ifx\csname bibfnamefont\endcsname\relax
  \def\bibfnamefont#1{#1}\fi
\expandafter\ifx\csname citenamefont\endcsname\relax
  \def\citenamefont#1{#1}\fi
\expandafter\ifx\csname url\endcsname\relax
  \def\url#1{\texttt{#1}}\fi
\expandafter\ifx\csname urlprefix\endcsname\relax\def\urlprefix{URL }\fi
\providecommand{\bibinfo}[2]{#2}
\providecommand{\eprint}[2][]{\url{#2}}

\bibitem[{\citenamefont{Wolf et~al.}(2001)\citenamefont{Wolf, Awschalom,
  Buhrman, Daughton, {von Moln{\'{a}}r}, Roukes, Chtchelkanova, and
  Treger}}]{Wolf2001}
\bibinfo{author}{\bibfnamefont{S.~A.} \bibnamefont{Wolf}},
  \bibinfo{author}{\bibfnamefont{D.~D.} \bibnamefont{Awschalom}},
  \bibinfo{author}{\bibfnamefont{R.~A.} \bibnamefont{Buhrman}},
  \bibinfo{author}{\bibfnamefont{J.~M.} \bibnamefont{Daughton}},
  \bibinfo{author}{\bibfnamefont{S.}~\bibnamefont{{von Moln{\'{a}}r}}},
  \bibinfo{author}{\bibfnamefont{M.~L.} \bibnamefont{Roukes}},
  \bibinfo{author}{\bibfnamefont{A.~Y.} \bibnamefont{Chtchelkanova}},
  \bibnamefont{and} \bibinfo{author}{\bibfnamefont{D.~M.}
  \bibnamefont{Treger}}, \bibinfo{journal}{Science}
  \textbf{\bibinfo{volume}{294}}, \bibinfo{pages}{1488} (\bibinfo{year}{2001}).

\bibitem[{\citenamefont{Sergueev et~al.}(2002)\citenamefont{Sergueev, Sun, Guo,
  Wang, and Wang}}]{PhysRevB.65.165303}
\bibinfo{author}{\bibfnamefont{N.}~\bibnamefont{Sergueev}},
  \bibinfo{author}{\bibfnamefont{Q.-F.} \bibnamefont{Sun}},
  \bibinfo{author}{\bibfnamefont{H.}~\bibnamefont{Guo}},
  \bibinfo{author}{\bibfnamefont{B.~G.} \bibnamefont{Wang}}, \bibnamefont{and}
  \bibinfo{author}{\bibfnamefont{J.}~\bibnamefont{Wang}},
  \bibinfo{journal}{Phys. Rev. B} \textbf{\bibinfo{volume}{65}},
  \bibinfo{pages}{165303} (\bibinfo{year}{2002}).

\bibitem[{\citenamefont{Pustilnik and Borda}(2006)}]{pustilnik:201301}
\bibinfo{author}{\bibfnamefont{M.}~\bibnamefont{Pustilnik}} \bibnamefont{and}
  \bibinfo{author}{\bibfnamefont{L.}~\bibnamefont{Borda}},
  \bibinfo{journal}{Phys. Rev. B} \textbf{\bibinfo{volume}{73}},
  \bibinfo{eid}{201301} (\bibinfo{year}{2006}).

\bibitem[{\citenamefont{Tao et~al.}(2008)\citenamefont{Tao, Shao-Quan, Ai-Hua,
  Fu-Bin, and Wei-Li}}]{Tao2008}
\bibinfo{author}{\bibfnamefont{H.}~\bibnamefont{Tao}},
  \bibinfo{author}{\bibfnamefont{W.}~\bibnamefont{Shao-Quan}},
  \bibinfo{author}{\bibfnamefont{B.}~\bibnamefont{Ai-Hua}},
  \bibinfo{author}{\bibfnamefont{Y.}~\bibnamefont{Fu-Bin}}, \bibnamefont{and}
  \bibinfo{author}{\bibfnamefont{S.}~\bibnamefont{Wei-Li}},
  \bibinfo{journal}{Chinese Phys. Lett.} \textbf{\bibinfo{volume}{25}},
  \bibinfo{pages}{2198} (\bibinfo{year}{2008}).

\bibitem[{\citenamefont{Goldhaber-Gordon
  et~al.}(1998)\citenamefont{Goldhaber-Gordon, Shtrikman, Mahalu,
  Abusch-Magder, Meirav, and Kastner}}]{Goldhaber-Gordon1998}
\bibinfo{author}{\bibfnamefont{D.}~\bibnamefont{Goldhaber-Gordon}},
  \bibinfo{author}{\bibfnamefont{H.}~\bibnamefont{Shtrikman}},
  \bibinfo{author}{\bibfnamefont{D.}~\bibnamefont{Mahalu}},
  \bibinfo{author}{\bibfnamefont{D.}~\bibnamefont{Abusch-Magder}},
  \bibinfo{author}{\bibfnamefont{U.}~\bibnamefont{Meirav}}, \bibnamefont{and}
  \bibinfo{author}{\bibfnamefont{M.~A.} \bibnamefont{Kastner}},
  \bibinfo{journal}{Nature} \textbf{\bibinfo{volume}{391}},
  \bibinfo{pages}{156} (\bibinfo{year}{1998}).

\bibitem[{\citenamefont{Cronenwett et~al.}(1998)\citenamefont{Cronenwett,
  Oosterkamp, and Kouwenhoven}}]{Cronenwett1998}
\bibinfo{author}{\bibfnamefont{S.~M.} \bibnamefont{Cronenwett}},
  \bibinfo{author}{\bibfnamefont{T.~H.} \bibnamefont{Oosterkamp}},
  \bibnamefont{and} \bibinfo{author}{\bibfnamefont{L.~P.}
  \bibnamefont{Kouwenhoven}}, \bibinfo{journal}{Science}
  \textbf{\bibinfo{volume}{281}}, \bibinfo{pages}{540} (\bibinfo{year}{1998}).

\bibitem[{\citenamefont{Schmid et~al.}(1998)\citenamefont{Schmid, Weis, Eberl,
  and {von~Klitzing}}}]{Schmid1998}
\bibinfo{author}{\bibfnamefont{J.}~\bibnamefont{Schmid}},
  \bibinfo{author}{\bibfnamefont{J.}~\bibnamefont{Weis}},
  \bibinfo{author}{\bibfnamefont{K.}~\bibnamefont{Eberl}}, \bibnamefont{and}
  \bibinfo{author}{\bibfnamefont{K.}~\bibnamefont{{von~Klitzing}}},
  \bibinfo{journal}{Physica B} \textbf{\bibinfo{volume}{256-258}},
  \bibinfo{pages}{182} (\bibinfo{year}{1998}).

\bibitem[{\citenamefont{van~der Wiel et~al.}(2002)\citenamefont{van~der Wiel,
  De~Franceschi, Elzerman, Fujisawa, Tarucha, and
  Kouwenhoven}}]{RevModPhys.75.1}
\bibinfo{author}{\bibfnamefont{W.~G.} \bibnamefont{van~der Wiel}},
  \bibinfo{author}{\bibfnamefont{S.}~\bibnamefont{De~Franceschi}},
  \bibinfo{author}{\bibfnamefont{J.~M.} \bibnamefont{Elzerman}},
  \bibinfo{author}{\bibfnamefont{T.}~\bibnamefont{Fujisawa}},
  \bibinfo{author}{\bibfnamefont{S.}~\bibnamefont{Tarucha}}, \bibnamefont{and}
  \bibinfo{author}{\bibfnamefont{L.~P.} \bibnamefont{Kouwenhoven}},
  \bibinfo{journal}{Rev. Mod. Phys.} \textbf{\bibinfo{volume}{75}},
  \bibinfo{pages}{1} (\bibinfo{year}{2002}).

\bibitem[{\citenamefont{Weidenm{\"{u}}ller}(2003)}]{PhysRevB.68.125326}
\bibinfo{author}{\bibfnamefont{H.~A.} \bibnamefont{Weidenm{\"{u}}ller}},
  \bibinfo{journal}{Phys. Rev. B} \textbf{\bibinfo{volume}{68}},
  \bibinfo{pages}{125326} (\bibinfo{year}{2003}).

\bibitem[{\citenamefont{Akera}(1993)}]{PhysRevB.47.6835}
\bibinfo{author}{\bibfnamefont{H.}~\bibnamefont{Akera}},
  \bibinfo{journal}{Phys. Rev. B} \textbf{\bibinfo{volume}{47}},
  \bibinfo{pages}{6835} (\bibinfo{year}{1993}).

\bibitem[{\citenamefont{Loss and Sukhorukov}(2000)}]{PhysRevLett.84.1035}
\bibinfo{author}{\bibfnamefont{D.}~\bibnamefont{Loss}} \bibnamefont{and}
  \bibinfo{author}{\bibfnamefont{E.~V.} \bibnamefont{Sukhorukov}},
  \bibinfo{journal}{Phys. Rev. Lett.} \textbf{\bibinfo{volume}{84}},
  \bibinfo{pages}{1035} (\bibinfo{year}{2000}).

\bibitem[{\citenamefont{Holleitner et~al.}(2001)\citenamefont{Holleitner,
  Decker, Qin, Eberl, and Blick}}]{PhysRevLett.87.256802}
\bibinfo{author}{\bibfnamefont{A.~W.} \bibnamefont{Holleitner}},
  \bibinfo{author}{\bibfnamefont{C.~R.} \bibnamefont{Decker}},
  \bibinfo{author}{\bibfnamefont{H.}~\bibnamefont{Qin}},
  \bibinfo{author}{\bibfnamefont{K.}~\bibnamefont{Eberl}}, \bibnamefont{and}
  \bibinfo{author}{\bibfnamefont{R.~H.} \bibnamefont{Blick}},
  \bibinfo{journal}{Phys. Rev. Lett.} \textbf{\bibinfo{volume}{87}},
  \bibinfo{pages}{256802} (\bibinfo{year}{2001}).

\bibitem[{\citenamefont{K\"{o}nig and Gefen}(2002)}]{PhysRevB.65.045316}
\bibinfo{author}{\bibfnamefont{J.}~\bibnamefont{K\"{o}nig}} \bibnamefont{and}
  \bibinfo{author}{\bibfnamefont{Y.}~\bibnamefont{Gefen}},
  \bibinfo{journal}{Phys. Rev. B} \textbf{\bibinfo{volume}{65}},
  \bibinfo{pages}{045316} (\bibinfo{year}{2002}).

\bibitem[{\citenamefont{Kubala and K{\"{o}}nig}(2002)}]{PhysRevB.65.245301}
\bibinfo{author}{\bibfnamefont{B.}~\bibnamefont{Kubala}} \bibnamefont{and}
  \bibinfo{author}{\bibfnamefont{J.}~\bibnamefont{K{\"{o}}nig}},
  \bibinfo{journal}{Phys. Rev. B} \textbf{\bibinfo{volume}{65}},
  \bibinfo{pages}{245301} (\bibinfo{year}{2002}).

\bibitem[{\citenamefont{B{\"{u}}ttiker
  et~al.}(1984)\citenamefont{B{\"{u}}ttiker, Imry, and Azbel}}]{Buttiker1984}
\bibinfo{author}{\bibfnamefont{M.}~\bibnamefont{B{\"{u}}ttiker}},
  \bibinfo{author}{\bibfnamefont{Y.}~\bibnamefont{Imry}}, \bibnamefont{and}
  \bibinfo{author}{\bibfnamefont{M.~Y.} \bibnamefont{Azbel}},
  \bibinfo{journal}{Phys. Rev. A} \textbf{\bibinfo{volume}{30}},
  \bibinfo{pages}{1982} (\bibinfo{year}{1984}).

\bibitem[{\citenamefont{Kubala and K{\"{o}}nig}(2003)}]{PhysRevB.67.205303}
\bibinfo{author}{\bibfnamefont{B.}~\bibnamefont{Kubala}} \bibnamefont{and}
  \bibinfo{author}{\bibfnamefont{J.}~\bibnamefont{K{\"{o}}nig}},
  \bibinfo{journal}{Phys. Rev. B} \textbf{\bibinfo{volume}{67}},
  \bibinfo{pages}{205303} (\bibinfo{year}{2003}).

\bibitem[{\citenamefont{Meir and Wingreen}(1992)}]{Meir1992}
\bibinfo{author}{\bibfnamefont{Y.}~\bibnamefont{Meir}} \bibnamefont{and}
  \bibinfo{author}{\bibfnamefont{N.~S.} \bibnamefont{Wingreen}},
  \bibinfo{journal}{Phys. Rev. Lett.} \textbf{\bibinfo{volume}{68}},
  \bibinfo{pages}{2512} (\bibinfo{year}{1992}).

\bibitem[{\citenamefont{Yamada}(1975{\natexlab{a}})}]{Yamada1975a}
\bibinfo{author}{\bibfnamefont{K.}~\bibnamefont{Yamada}},
  \bibinfo{journal}{Prog.~Theor.~Phys.} \textbf{\bibinfo{volume}{53}},
  \bibinfo{pages}{970} (\bibinfo{year}{1975}{\natexlab{a}}).

\bibitem[{\citenamefont{Yamada}(1975{\natexlab{b}})}]{Yamada1975}
\bibinfo{author}{\bibfnamefont{K.}~\bibnamefont{Yamada}},
  \bibinfo{journal}{Prog.~Theor.~Phys.} \textbf{\bibinfo{volume}{54}},
  \bibinfo{pages}{316} (\bibinfo{year}{1975}{\natexlab{b}}).

\bibitem[{\citenamefont{Yosida and Yamada}(1975)}]{Yosida1975}
\bibinfo{author}{\bibfnamefont{K.}~\bibnamefont{Yosida}} \bibnamefont{and}
  \bibinfo{author}{\bibfnamefont{K.}~\bibnamefont{Yamada}},
  \bibinfo{journal}{Prog.~Theor.~Phys.} \textbf{\bibinfo{volume}{53}},
  \bibinfo{pages}{1286} (\bibinfo{year}{1975}).

\bibitem[{\citenamefont{Yosida and Yamada}(1970)}]{Yosida1970}
\bibinfo{author}{\bibfnamefont{K.}~\bibnamefont{Yosida}} \bibnamefont{and}
  \bibinfo{author}{\bibfnamefont{K.}~\bibnamefont{Yamada}},
  \bibinfo{journal}{Prog. Theor. Phys. Supp.} \textbf{\bibinfo{volume}{46}},
  \bibinfo{pages}{244} (\bibinfo{year}{1970}).

\bibitem[{\citenamefont{Zlati{\'{c}} and Horvati{\'{c}}}(1980)}]{Zlati'c1980}
\bibinfo{author}{\bibfnamefont{V.}~\bibnamefont{Zlati{\'{c}}}}
  \bibnamefont{and}
  \bibinfo{author}{\bibfnamefont{B.}~\bibnamefont{Horvati{\'{c}}}},
  \bibinfo{journal}{Phys.~stat.~sol.} \textbf{\bibinfo{volume}{99}},
  \bibinfo{pages}{251} (\bibinfo{year}{1980}).

\bibitem[{\citenamefont{Zlati{\'{c}} and Horvati{\'{c}}}(1983)}]{Zlati'c1983}
\bibinfo{author}{\bibfnamefont{V.}~\bibnamefont{Zlati{\'{c}}}}
  \bibnamefont{and}
  \bibinfo{author}{\bibfnamefont{B.}~\bibnamefont{Horvati{\'{c}}}},
  \bibinfo{journal}{Phys. Rev. B} \textbf{\bibinfo{volume}{28}},
  \bibinfo{pages}{6904} (\bibinfo{year}{1983}).

\bibitem[{\citenamefont{Oguri}(2001)}]{Oguri2001a}
\bibinfo{author}{\bibfnamefont{A.}~\bibnamefont{Oguri}},
  \bibinfo{journal}{Phys. Rev. B} \textbf{\bibinfo{volume}{64}},
  \bibinfo{pages}{153305} (\bibinfo{year}{2001}).

\bibitem[{\citenamefont{Zlati\ifmmode~\acute{c}\else \'{c}\fi{} and
  Horvati\'{c}}(1983)}]{Zlatiifmmodeacutecelse'cfi1983}
\bibinfo{author}{\bibfnamefont{V.}~\bibnamefont{Zlati\ifmmode~\acute{c}\else
  \'{c}\fi{}}} \bibnamefont{and}
  \bibinfo{author}{\bibfnamefont{B.}~\bibnamefont{Horvati\'{c}}},
  \bibinfo{journal}{Phys. Rev. B} \textbf{\bibinfo{volume}{28}},
  \bibinfo{pages}{6904} (\bibinfo{year}{1983}).

\bibitem[{\citenamefont{Wilhelm et~al.}(2002)\citenamefont{Wilhelm, Schmid,
  Weis, and {von Klitzing}}}]{Wilhelm2002385}
\bibinfo{author}{\bibfnamefont{U.}~\bibnamefont{Wilhelm}},
  \bibinfo{author}{\bibfnamefont{J.}~\bibnamefont{Schmid}},
  \bibinfo{author}{\bibfnamefont{J.}~\bibnamefont{Weis}}, \bibnamefont{and}
  \bibinfo{author}{\bibfnamefont{K.}~\bibnamefont{{von Klitzing}}},
  \bibinfo{journal}{Physica E} \textbf{\bibinfo{volume}{14}},
  \bibinfo{pages}{385 } (\bibinfo{year}{2002}).

\end{thebibliography}

\end{document}